# Mapping the Digital Healthcare Revolution*

**Marcelo Corrales Compagnucci, Mark Fenwick, Michael Lowery Wilson, Nikolaus Forgó and Till Bärnighausen**

**Abstract:** This introductory chapter briefly outlines the main theme of this volume, namely, to review the new opportunities and risks of digital healthcare from various disciplinary perspectives. These perspectives include law, public policy, organisational studies, and applied ethics. Based on this inter-disciplinary approach, we hope that effective strategies may arise to ensure that benefits of this on-going revolution are deployed in a responsible and sustainable manner. The second part of the chapter comprises a brief review of the four parts and fourteen substantive chapters that comprise this volume.

**Keywords:** AI, eHealth, privacy, data protection, consent, GDPR, personalised medicine, risks, trust, ethics, platforms, hospitals, medical data, genetic data, AI wearables

## 1 Introduction

Digital technologies are disrupting health care and creating new opportunities and risks for all actors in the medical ecosystem. Moreover, many of these developments rely heavily on data and AI algorithms to prevent, diagnose, treat, and monitor sources of epidemic diseases, such as the ongoing COVID-19 pandemic and other pathogenic outbreaks. However, these opportunities and risks have a complex character involving multiple dimensions (notably legal, ethical, technical and

---

governance) and any mapping and navigation of this new space requires an appreciation of the complexity of these issues and multidisciplinary dialogue.

This introductory chapter briefly outlines the main theme of this volume, namely, to review the new opportunities and risks of digital healthcare from various disciplinary perspectives – specifically law, public policy, organisational studies, and applied ethics. Based on this inter-disciplinary approach, we hope that effective strategies to ensure that the benefits of this on-going revolution are deployed in a responsible and sustainable way can be developed. The second part of the chapter consists of an overview of the four constituent parts and fourteen substantive chapters that comprise this volume.

## 2 Challenges and Strategies of AI in Digital Healthcare

In recent years, there has been an increasing awareness of the vital role that AI plays in various domains of economic and social life to resolve multiple issues. By way of a provisional definition, AI aims at simulating human intelligence (for example, by planning, strategizing and making advanced decisions).[1] Particularly significant in this regard, AI systems are being developed to analyse the massive amounts of medical and genetic data, understand human conditions, recognise disease patterns, make highly accurate diagnoses, and deliver precision health interventions at scale.[2] There are several kinds of AI tools and techniques currently being utilised across a number of settings including hospitals, clinical laboratories, and research facilities.[3] For this reason, AI is widely predicted to provide the foundations for the "next Industrial Revolution" and provide the driving force that will disrupt how healthcare is delivered and experienced in the future.[4]

Digital health can be understood as the convergence of digital technologies with health, healthcare, living, and society to enhance the efficiency of healthcare delivery and make medicine more personalised and effective. The broad scope of digital health includes categories such as mobile health, health information technology, wearable devices, tele-health and telemedicine, and personalised medicine. These technologies can empower patient-consumers to make better-informed decisions about their own health and provide new options for facilitating prevention, early diagnosis of life-threatening diseases, and management of chronic conditions outside of traditional care settings.[5]

---

[1] Gai-zhen Yang, 'Office Operating Problem Scoring System Based on AI' in: Hui YANG (ed) *Artificial Intelligence: Science and Technology*, Proceedings of the 2016 International Conference (AIST 2016), Shanghai, China (World Scientific, 2017), 21.

[2] Arvin Agah, *Medical Applications of Artificial Intelligence* (CRC Press, 2014).

[3] Arjun Panesar, *Machine Learning and AI for Healthcare: Big Data for Improved Health Outcomes* (Apress, 2019).

[4] Prisilla Jayanthi et al., 'Fourth Industrial Revolution: An Impact on Healthcare Industry' in: Tareq Ahram (ed) *Advances in Artificial Intelligence, Software and Systems Engineering* (Springer, 2019), 58.

[5] Letizia Afinito, *Empowering the Connected Physician in the E-Patient Era: How Physicians Empowerment on Digital Health Tools Can Improve Patient Empowerment and Boost Health(care) Outcomes* (Routledge, 2019).

A lot of developments fall within the scope of this definition. From mobile medical apps and software that support the clinical decisions doctors make every day, to artificial intelligence (AI) and machine learning (ML), digital technology is driving a revolution in health care. Digital health tools have the potential to improve our ability to accurately diagnose and treat disease and to enhance the delivery of health care for the individual. Digital tools are also offering health care providers a more holistic view of patient health through access to data and giving patients more control over their health. Digital health offers genuine opportunities to improve medical outcomes and enhance efficiency.

At the level of populations, more granular data and new technologies have driven the growth of what is now called precision public health (PPH), providing governments and private companies with new mechanisms for offering more effective interventions. Again, PPH is intimately connected with developments in AI as it leverages data and predictive analytics to identify health risks, detect diseases more rapidly, and design interventions for subpopulations that reach the appropriate target audience.[6] PPH also holds out the tantalising possibility of more effective prevention and individualised interventions at lower costs and delivering better healthcare to individuals in low-income environments who lack insurance or access to facility-based healthcare.[7] PPH begs the questions how individual-focused care approaches can be reconciled with benefits on a population scale – in a manner that respects the individual, ensures privacy and increases, rather than decreases, autonomy and choice.[8] As in other domains of healthcare, this requires a combination of ethical principles, a robust and multi-faceted regulatory framework, and robust governance structures.

While opening a world of new opportunities, however, rapid advances in AI have been compared to a "black box," potentially unleashing several serious ethical dilemmas and raising uncertainty about the current legal framework on privacy and data protection. It has been argued that AI systems, for example, may run afoul of the consent of data subjects as such systems often collect, process, and transfer sensitive personal data in unexpected ways without the necessary means of giving adequate notice, choice, and explaining options in a timely manner.[9]

Despite promising results, the application of AI in medical devices must still confront technological, legal and ethical issues.[10] A serious limitation lies in the lack of interoperability and standardisation among medical IT systems,[11] and healthcare provision often involves complex judgments and abilities that AI is currently unable to replicate, such as the ability to read social cues.[12] Since AI medical devices can err, reliability and safety are crucial issues, particularly in the early stages of development when awareness of knowledge of likely problems is much less developed.[13]

---

[6] Adam Dunn et al., 'Social Media Interventions for Precision Public Health: Promises and Risks, *NPG Digital Medicine* (2018)1:47.

[7] Shawn Dolley, 'Big Data's Role in Precision Public Health' *Frontiers in Public Health*, March (2018)6:68.

[8] Mattia Prosperi, 'Big data hurdles in precision medicine and precision public health' *BMC Medical Informatics and Decision Making* (2018)18:139.

[9] Woodrow Barfield and Ugo Pagallo (eds) *Research Handbook on the Law of Artificial Intelligence* (Edward Elgar Publishing, 2018), 280-385.

[10] Sara Gerke, Serena Young & Glenn Cohen, 'Ethical and Legal Aspects of Ambient Intelligence in Hospitals' *JAMA* (Jan. 24, 2020).

[11] G Brindha, 'A New Approach for Changes in Health Care' *Middle-East Journal of Scientific Research* (2012), 12(12):16571662.

As such, these complex technological developments raise several important and difficult questions. What impact will AI systems have on biomedical and automated scientific research, especially on data sharing and confidentiality? What kind of control over personal data should be adjudicated to patients? How can we ensure that AI-based methods and solutions adhere to general legal and ethical principles? And how will these technological advancements in the MedTech industry be affected by different legal frameworks? How can we ensure that AI-based methods and solutions adhere to general legal and ethical principles? Which regulatory, ethical, and legal principles should guide the design of precision public health interventions and the implementation of precision medicine?

Regulators and other policymakers are reacting to these technological and professional issues with several important initiatives. The GDPR has tried to adequately respond to some of these challenges, for example by its rule on automated decision making. A striking feature of the GDPR is its potentially global reach and this might have prompted legislators to carry out reforms in other jurisdictions outside of the EU. However, many uncertainties and lingering questions still remain regarding the scope, direction and effects of the impact of AI in digital health care systems and personalised medicine. For instance, a serious problem of the GDPR is that the regulation, due to the high level of abstraction it adopts, is not capable of adequately differentiating between different applications of AI in a medical context. Addressing the many challenges generated by AI, therefore, requires going beyond any one disciplinary perspective or frame of reference. This means that we need a more seamlessly integrated or interdisciplinary approach as there are still multiple concerns to be resolved.

It is instructive in this regard to focus on the GDPR as an example. There are several provisions within the GDPR that allow for the processing of health data for scientific research to for example inform population health decision-making. The European Data Protection Board (EDPB) on 21 April 2020, published the Guidelines 03/2020 on the processing of data concerning health for the purpose of scientific research in the context of the COVID-19 outbreak.[14] The EDPB guidelines support research and data sharing under the appropriate legal framework. For example, data which is transmitted by devices and applications should include both unique and pseudonymous identifiers. These identifiers should be generated by the application and be specific to it. They also have to be renewed on a regular basis at intervals which are compatible with the goals of containing the virus spread. These aspects should also be done in a manner that allows for a reduction in the risk of identification or tracking of individual persons. However, even in these cases, the EDPB states that any data processing must be transparent, and that the data should be processed with sufficient privacy safeguards in place and not shared with third parties without prior authorisation.

[12] Max Louwerse et al., 'Social Cues in Animated Conversational Agents' *Applied Cognitive Psychology: The Official Journal of the Society for Applied Research in Memory and Cognition* (2005), 19(6):693704.

[13] Robert M. Wachter, 'Why Diagnostic Errors Don't Get Any Respect and What Can Be Done About Them' *Health Affairs* (2017), 29(9) https://doi.org/10.1377/hlthaff.2009.0513 (accessed 27 July 2021).

[14] Guidelines 03/2020 on the processing of data concerning health for the purpose of scientific research in the context of COVID-19, available at: https://edpb.europa.eu/our-work-tools/our-documents/ohjeet/guidelines-032020-processing-data-concerning-health-purpose_en (accessed 20 July 2021).

Another important strategy pursued by the EU has been the release of guidelines[15] to encourage the development of trustworthy and more ethical AI. The Ethics Guidelines for Trustworthy AI were published on 8 April 2019 by the High-Level Expert Group on AI and they received more than 500 comments after open consultation.[16] Although not legally binding, they address some of the diffuse problems that AI will bring to society as we integrate it in sectors such as: healthcare, education, and consumer technology. The Guidelines focus on how governments, companies, and other organisations need to develop ethical applications of AI. According to the Guidelines, AI systems should be accountable, explainable, and unbiased. To help achieve this goal, the EU recommends using an assessment list of seven fundamental areas that AI systems should meet in order to be deemed trustworthy. Among these requirements, human autonomy, privacy and data governance are at the core. Personal data collected by AI systems should be lawful, secure, resilient, reliable, robust and private. The Guidelines also underscore the importance of 'transparency.' Data and algorithms used to create AI systems should be accessible and traceability should be ensured. Similar guidelines can be found in other jurisdictions and the use of guidelines is a striking feature of the contemporary regulatory landscape regarding AI.

More traditional legal forms are still relevant. On 21 April 2021, the European Commission released its draft regulation governing the use of AI. The proposed AI Regulation[17] follows a risk-based approach with different categories of AI system uses such as prohibited, high-risk, limited and minimal risk. Prohibited AI systems are those which contravene union values (eg, by violating fundamental rights) and are considered unacceptable. The high-risk category will be subject to stricter regulatory requirements before and after releasing the product to the market (eg, ensuring the quality of data sets used to train the algorithm, applying a level of human agency and oversight, providing relevant information to users, etc). Nevertheless, makers of 'limited' or 'minimal' risk (eg, where there is a risk of manipulation, for example via the use of chatbots), will be encouraged to adopt non-legally binding codes of conduct.[18]

The GDPR and the above-mentioned AI draft regulation/guidelines are just a few examples that have, inevitably given the importance of the EU, received a lot of media and academic attention, but policy initiatives are occurring across the globe. Several other international organisations have published guidance on AI such as the OECD Council Recommendation on Artificial Intelligence[19] which promotes AI that is innovative and trustworthy and that respect human rights and democratic values. The OECD Council Recommendation on AI is the first of such principles signed up by governments not part of the OECD such as Argentina, Brazil, Costa Rica, Peru, Malta, Romania and Ukraine. Another example is the recently published World Health Organisation (WHO) guidance on the Ethics and Governance of AI for Health.[20] This guidance is based on six principles: Protecting human autonomy; promoting human well-being and safety and the public interest; ensuring transparency, explainability and intelligibility; fostering responsibility and accountability; ensuring inclusiveness and equity; promoting AI that is responsive and sustainable. This is clearly a fast-moving space, but the basis for future regulation has already started to emerge and disseminating information and subjecting these developments to rigorous review seems important, as initiatives taken now seem likely to structure debate and regulatory responses for the foreseeable future.

---

[15] European Commission, Ethics Guidelines for Trustworthy AI, available at: *https://ec.europa.eu/digital-single-market/en/news/ethics-guidelines-trustworthy-ai* (accessed 27 July 2021).

[16] European Commission, Shaping Europe's Digital Future, Ethics Guidelines for Trustworthy AI, Report Study, available at: https://digital-strategy.ec.europa.eu/en/library/ethics-guidelines-trustworthy-ai (accessed 27 July 2021).

[17] At the time of writing, the EU Member States have not yet adopted the proposed AI Regulations.

[18] Julia Wilson, New Draft Rules on the Use of Artificial Intelligence (14 May 2021), available at: https://www.bakermckenzie.com/en/insight/publications/2021/05/new-draft-rules-on-the-use-of-ai (Accessed 27 July 2021).

## 3 Argumentation & Structure

This edited collection brings together a series of contributions by leading scholars from different disciplines and diverse nationalities to examine the technical features that are driving the development of AI in medical contexts, as well as the efficacy of the current regulatory responses. As such, this book offers a high-level, global, and interdisciplinary perspective on current debates on AI in eHealth. The book attempts to navigate the contours of the highly complex ethical dilemmas and legal challenges raised by these disruptive technologies with the aim of designing practical proposals.

The unique selling point of this collection is the international and multidisciplinary background of its contributors. Represented disciplines include medicine and law, but also management, philosophy, and computer science. Chapters are tied together by a focus on bringing all these disciplines and their associated policy proposals into better alignment and deepening our understanding of the various responses to these game-changing technological, economic, legislative, and social developments.

The book comprises fourteen thematic chapters divided into four main parts (Part I – 'Platforms, Apps & Digital Health'; Part II – 'Trust & Design'; Part III – 'Knowledge, Risk & Control'; Part IV – 'Balancing Regulation, Innovation & Ethics'). Each part focuses on different technical, legal, and ethical processes and outcomes, providing stimulation for beginners and experts, academia, and business. It is our hope that this collection illustrates the art of emerging possibilities across the many levels and dimensions which lie at the interface between AI and

---

[19] OECD Recommendation of the Council on Artificial Intelligence, adopted in May 2019 by the OECD, available at: https://legalinstruments.oecd.org/en/instruments/OECD-LEGAL-0449

[20] Ethics and Governance of Artificial Intelligence for Health, WHO Guidance (28 June 2021), available at: https://www.who.int/publications/i/item/9789240029200 (accessed 27 July 2021).

eHealth.

## Part I: Platforms, Apps & Digital Health

The three chapters in Part I explore the impact of software applications – often developed by companies with software, rather than medical expertise in the digital healthcare space.

Chapter Two, 'Technology-Driven Disruption of Healthcare & UI Layer Privacy-by-Design' by Marcelo Corrales Compagnucci, Mark Fenwick, Helena Haapio, Timo Minssen and Erik P. M. Vermeulen describes how the use of digital technologies in healthcare is changing how medical treatments are developed by researchers, practised by medical professionals, and experienced by patients. The chapter argues that a defining feature of this disruption is the emergence of new medical 'apps' that leverage algorithm-based AI systems. As the use of such apps and AI wearables goes mainstream and new players – notably 'Super Platforms' with digital rather than a medical expertise – enter the healthcare sector, the traditional means of providing medical services are further disrupted.

These developments pose several challenges for regulators and other policymakers, most obviously, in the areas of privacy and data protection. The chapter describes how the emerging field of Legal Design can provide a more transparent and accessible infrastructure that embeds relevant legal protections 'in' the user interfaces of healthcare products and services. This user interface focused Privacy-by-Design approach offers multiple advantages, most obviously greater transparency, accountability, and choice. The chapter offers several real-world examples of design patterns that illustrate the value of UI focused Privacy-by-Design in protecting individuals' sensitive information, enabling people to make choices and retain control of their personal data. The chapter concludes by reflecting on the challenges specific to implementing Legal Design in an eHealth context.

In Chapter Three, 'Social Media Platforms as Public Health Arbiters: Global Ethical Considerations on Privacy, Legal and Cultural Issues Associated with Suicide Detection Algorithms,' by Karen L. Celedonia, Michael Lowery Wilson and Marcelo Corrales Compagnucci, the authors discuss the issue of the responsibility of social media firms for medical issues. The development of Facebook's suicide prevention algorithm has prompted discussion around whether social media platforms have a role to play in public health surveillance. Concerns have been raised about an entity that is not a public interest health authority collecting and acting on the private health information of its users, particularly when it involves personally sensitive data, such as an individual's mental health status. Mental illnesses are still heavily stigmatised, despite continued efforts to normalise these conditions. Depending on a user's geographic location, the ramifications of the suicide detection algorithm generating 'false positives' for suicide risk could have severe repercussions. This chapter seeks to stimulate further debates on this question by examining the ethical implications of Facebook's suicide prevention algorithm from diverse perspectives.

In Chapter Four, 'Promoting the Use of PHR by Citizens and Physicians – Proposed Design for a Token to be Allocated to Citizens,' Shinto Teramoto focuses on health records. The digitalised medical and health records of citizens are stored in the Electronic Health Records (EHR) of hospitals or clinics, and in Personal Healthcare Records (PHR). The quality of medical care is improved if physicians have access to the complete past records of patients. A user-friendly service that enables individual citizens to share their health and medical records in PHR with their physicians is, therefore, essential to achieving this objective. To encourage patients and physicians to share medical records utilising PHR, while avoiding conflict with the recent trend demanding that citizens have autonomous control of their own personal information, governments must develop various legal measures to encourage individual citizens to take the initiative to record their medical and health data in their PHR and to give their physicians access to PHR. The chapter proposes mathematical schemes that might be implemented within the framework of the existing regulatory framework.

## Part II: Trust & Design

Part II consists of three chapters looking closely at data protection issues with a particular emphasis on questions of consent and trust.

Chapter Five, 'Privacy Management in eHealth Using Contextual Consenting' by Yki Kortesniemi and Päivi Pöyry-Lassila starts from the fact that sharing one's health data with one's doctor can be an important factor in improving one's own health and sharing data for scientific research can help improve the health of everyone. At the same time, health data is highly confidential, so the sharing process must provide sufficient control over one's privacy. Legally, sharing is often based on consent, which theoretically affords extensive individual control, but in practice often requires the processing of complicated information. Therefore, the way the consenting process is implemented plays a significant role in either hindering or helping the individual. This chapter illustrates the potential of AI-based technologies and explores how an individual's ability to make informed consenting decisions can be simplified by utilising AI-based recommendations with the consent intermediary approach and by making the consenting decisions in the context of utilising the health data thus making individuals more aware of the data they are sharing.

In Chapter Six, 'Artificial Intelligence and Data Protection Law,' Thomas Hoeren and Maurice Niehoff describe how the increasing automation of medical decision-making is also accompanied by a range of new problems, in particular the maintenance of the relationship of trust between physicians and patients or the verification of decisions. This is where the patient's right to explanation comes into play, which is enshrined in the GDPR. This chapter explains how the right is derived from the GDPR and how it should be established in the context of automated medical decision-making.

Chapter Seven, 'AI Technologies and Accountability in Digital Health' by Eva Thelisson

focuses on a similar question, namely, how to build an ecosystem of trust in this new arena of digital health? The availability of large amounts of personal data, from multimodal sources, combined with AI and ML capacities, Internet of Things and strong computational platforms have the potential to transform healthcare systems in a disruptive way. The emergence of personalised medicine offers opportunities and raises new legal, ethical, and societal challenges. A silent transformation towards a data-driven preventive and personalised medicine may improve diagnosis and therapies while reducing the cost of public health policy. In order to build an ecosystem of trust, the risks of harm and misuses such as data breaches, privacy issues, discrimination, eugenics must be addressed. This chapter presents the disruptive nature of AI and ML technologies in healthcare and makes specific recommendations to build a trustworthy digital health system. The chapter first identifies some general parameters to advance the field of digital health in a responsible way, and, secondly, proposes possible solutions to shape a sound policy in digital health taking into consideration a rights-based governance framework.

## Part III: Knowledge, Risk & Control

Part III, comprising three chapters, explores various risks that arise as a result of the emergence of new forms of knowledge produced by AI-related analysis of medical data.

Chapter Eight, 'The Principle of Transparency in Medical Research: Applying Big Data Analytics to Electronic Health Records' by Nikolaus Forgó and Marie-Catherine Wagner describes how in recent years, the amount of data provided by EHRs worldwide has greatly expanded bringing obvious benefits to diverse stakeholders. The more heath data that is collected, the more can be learned from it and better decisions can be made based on Big Data analysis of that data. This can be seen in projects such as the InteropEHRate project, an EU Horizon 2020 project, which tries to provide models on how health data from EHRs can be made interoperable and available for medical research organisations. However, the processing of personal data in this way might interfere with the fundamental right to data protection or privacy. On a European level, the GDPR treats specific forms of data processing differently, if it is to be expected that those are specifically protected by other, potentially conflicting fundamental rights and freedoms. The GDPR provides privileges for scientific research in some respects and allows additional derogations for Member States. In particular, Art 89 (2) GDPR provides exemptions from data subjects' rights. When health data are analysed on the basis of ML, special attention needs to be paid to the transparency principle, which is a fundamental feature in EU law as – evidently – transparency is both needed and challenged when machines, replacing or supporting humans, take decisions. This chapter provides an analysis of the principle of transparency and its compatibility with Big Data analytics in medical research. Apart from an evaluation of the current European legal framework, including the Council of Europe's Convention 108+, the chapter also examines global initiatives, such as the 'Recommendation on the Protection and Use of Health-Related Data', whose final text was presented by the UN Special Rapporteur for Privacy to the UN General Assembly in October 2019.

Chapter Nine, 'The Next Challenge for Data Protection Law: AI Revolution in Automated Scientific Research' by Janos Meszaros proceeds from the observation that although an extensive literature has been published on autonomous vehicles, robotics in healthcare, and the disruption of work by automation, there has been relatively little discussion on how AI might change scientific research itself. AI-assisted scientific research is already providing a significant boost in the process of scientific discovery, particularly in a medical context. Not surprisingly, this radical change in scientific research will have significant consequences. Firstly, if the research process becomes automated, it may be conducted by anyone, which puts citizen science in a new context. As developments in hardware and software made personal computers feasible for individual use, automated research may have a similar effect on science in the future. Secondly, unlike researchers, AI and neural networks cannot explain their thinking yet. As fully automated research extends the potential "black box" of AI even further, this makes the oversight and ethical review more problematic in systems that are opaque to outside scrutiny. Automated research raises many further questions about regulation, safety, funding, and patentability. This chapter will focus on the issues connected with privacy and data protection, from the GDPR point of view.

In Chapter Ten, 'A Global Human-Rights Approach to Medical Artificial Intelligence,' by Audrey Lebret the focus is on the role of algorithms. The use and development of algorithms in health care, including ML, contributes to the discovery of better treatments for patients and offers promising perspectives in the fight against cancer and other diseases. Yet, algorithms are not a neutral health product since they are programmed by humans, with the risk of propagating human rights infringements and discrimination. In the medical area, human rights impact assessments need to be conducted for applications involving AI. Apart from offering a consistent and transversal substantive approach to AI, human rights law, and in particular the UN guiding principles on business and human rights, would allow the targeting of all stakeholders, including the corporations developing health care algorithms. Such an approach would establish a chain of duties and responsibilities bringing more transparency and consistency in the overall AI development process and later uses. Although this approach may not solve all AI challenges, it could offer a frame for discussion with all relevant actors, including vulnerable populations. An increase in human rights education of medical doctors and data scientists, and further collaboration at the initial stages of the development of algorithms would greatly contribute to the creation of a human rights culture in this fast-developing techno-science space.

## Part IV: Balancing Regulation, Innovation & Ethics

Part IV, comprising four chapters, examines the challenges of balancing the different concerns that arise in real world settings, most obviously in hospitals and in physician-patient relations.

In Chapter Eleven, 'Doctors without Borders? The Law Applicable to Cross-Border eHealth Services and AI-based Medicine,' Jan D. Lüttringhaus proceeds from the idea that health applications – including telemedicine, AI-based medicine and smart medical devices – are ubiquitous. Such tools may be used by the physician located next door as well as in the most remote locations abroad. Moreover, highly sensitive medical data may flow around the world within a split second. Against this backdrop, eHealth and telemedicine services can be provided from – and the necessary data can be transferred to – virtually every corner of the world. By contrast, the scope of application of regulation relating to AI-driven medicine as well as eHealth- and telemedicine is usually confined to the legislating state. Moreover, the number and complexity of rules and regulations in this field varies considerably from state to state. Does this mean that international 'MedTech'-businesses may simply set up camp in the jurisdiction most favourable to their business models? For practitioners in telemedicine, the MedTech-industry providing AI applications or digital medical devices such as eHealth-apps as well as for patients, it is essential to know which law governs activities are undertaken in cross-border scenarios: This concerns licensing requirements and the level of data protection as well as contract and tort law applicable to eHealth, telemedicine and telesurgery services.

Chapter Twelve, 'Barriers to Artificial Intelligence in Hospitals and Arguments for Developing a Hospital-Specific AI Readiness Index' by Maximilian Schuessler, Till Bärnighausen and Anant Jani describes how AI has considerable potential to improve diagnosis and therapy, enhance access to healthcare, and promote population health. Although in its infancy, AI-enabled healthcare is increasingly seen as part of the solutions needed to address the growing gap between the supply and demand of hospital care. AI is well placed to help us tackle new challenges, though these novel applications are likely to render technology implementation even more complex. AI technologies are on the cusp of entering hospitals. Yet, many hospitals within the EU are unprepared for this change. Historically, hospitals have faced multiple challenges when implementing new technologies. This chapter discusses the importance of AI readiness and highlights the benefits and limitations of a new policy tool: an AI Readiness Index for Hospitals (AI-RIH). The authors conceptualise AI readiness from an organisational perspective and discuss the dual functionality of the AI-RIH. For hospital managers, such an index could constitute a benchmarking tool. For policy-makers, national and EU-wide, knowledge about AI readiness and changes therein can help customise targeted technology policies and measure their effectiveness. The chapter also discusses conceptual challenges of indices and illustrates why a hospital index might provide more policy insights than an aggregated or national index. Finally, it explains how AI readiness can strengthen hospitals' role as innovators and support the development and deployment of AI.

Marc Stauch in Chapter Thirteen, 'Regulating the Benefits of eHealth – Information Disclosure Duties in the Age of AI,' looks at how much of the legal and ethical attention in the fields of eHealth focuses on the risks of health data processing 'going wrong' – a breach of privacy occurs, data is misused in an unauthorised way, or the analysis of data gives a faulty result. However, significant challenges are also posed by such processing where the data processing 'goes to plan' – the analysis gives the correct result in the way intended. Such challenges stem both from the nature of the information generated, and the new contexts in which this occurs. Thus, Big Data analysis may produce ever more information in relation to a person's future health, usually of a probabilistic nature. In what situations should such information be returned to the subject (bearing in mind also that the decision-maker increasingly will be an entity outside the traditional health care context)? This chapter considers key ethical considerations that arise in such cases, and how well the law – through liability rules for harm, caused by failure to disclose, or by unjustified disclosure – is equipped to respond to these complex situations.

Chapter Fourteen, 'Privacy and Direct-to-Consumer Genetic Tests,' by Dena Dervanović examines the growing interest of law enforcement authorities' in using DTC genetic test providers' databases for solving crime. The chapter discusses the legal avenues that were used by the Swedish police authority in their use of GEDmatch to resolve a 16-year-old double murder. It discusses the legal prerequisites for genetic test data access and use as well as embarks on a discussion about the possibility of relying on the derogation of special categories of personal data which are made public by the data subject. The chapter also discusses possible amendments to the existing legal landscape around such data.

In Chapter Fifteen 'Health Research, eHealth and Learning Healthcare Systems: Key Approaches, Shortcomings and Design Issues in Data Governance,' Shawn Harmon examines how the pressure to collect more health data and use that data more effectively is mounting as healthcare systems face greater challenges. However, the risks of increasing health data collection and making our health data work harder are myriad. Given that 'good outcomes' in relation to health data usage will be context specific and temporally contingent, the emphasis here is on fit-for-purpose instruments and good practice, acknowledging that health data usage is mediated not only through law, but also through governance structures around data resources themselves. This chapter therefore reviews the Canadian health data ecosystem, examining its federal and provincial legislative elements (with an emphasis on Nova Scotia). It then critiques that ecosystem, bearing in mind the needs of learning healthcare systems. In doing so, it highlights four ecosystem shortcomings, which are grounded in no small part on the perceived competition between private and public interests, and the poor alignment between contemporary data uses and traditional protections associated with autonomy (consent) and privacy (anonymisation). Finally, it offers some key considerations for ecosystem design, addressing specifically social licenses to operate and the value foundation of both legislation and repository governance instruments.

Our primary intention in putting together this collection is to stimulate further debate on the various issues raised and to provide a framework for thinking about effective strategies to ensure that the benefits of this on-going health care revolution are developed in a responsible and sustainable way.

# Bibliography

Afinito L, Empowering the Connected Physician in the E-Patient Era: How Physicians Empowerment on Digital Health Tools Can Improve Patient Empowerment and Boost Health(care) Outcomes (Routledge, 2019).

Agah A, (2014) Medical Applications of Artificial Intelligence. CRC Press, Boca Ratón;

Badia, RM, Corrales, M, Dimitrakos, T, Djemame, K, Elmroth, E, Ferrer, AJ, Forgó, N, Guitart, J, Hernández, F, Hudzia, B, Kipp, A, Konstanteli, K, Kousiouris, G, Nair, SK, Sharif, T, Sheridan, C, Sirvent, R, Tordsson, J, Varvarigou, T, Wesner, S, Ziegler, W & Zsigri, C 2011, 'Demonstration of the OPTIMIS Toolkit for Cloud Service Provisioning', *Lecture Notes in Computer Science (including subseries Lecture Notes in Artificial Intelligence and Lecture Notes in Bioinformatics)*, pp. 331-333. https://doi.org/10.1007/978-3-642-24755-2_40

Barfield W and Pagallo U (eds) Research Handbook on the Law of Artificial Intelligence, (Edward Elgar Publishing, 2018).

Brindha G, A new approach for changes in health care. Middle-East Journal of Scientific Research, 12(12):16571662, 2012.

Celedonia, KL, Corrales Compagnucci, M, Minssen, T & Lowery Wilson, M 2021, 'Legal, Ethical, and Wider Implications of Suicide Risk Detection Systems in Social Media Platforms', *Journal of Law and the Biosciences*, vol. 8, no. 1. https://doi.org/10.1093/jlb/lsab021

Celedonia, K, Valenti, M, Corrales Compagnucci, M & Lowery Wilson, M 2021, 'Community-Based Health Care Providers as Research Subject Recruitment Gatekeepers: Ethical and Legal Issues In A Real-World Case Example', *Research Ethics*, vol. 17, no. 2, pp. 242-250. https://doi.org/10.1177/1747016120980560

Chowkwanyn M, "Precision" Public Health — Between Novelty and Hype, New England Journal of Medicine, 379;15 nejm.org October 11, 2018.

Corrales Compagnucci, M, Aboy, M & Minssen, T 2021, 'Cross-Border Transfers of Personal Data after Schrems II: Supplementary Measures and New Standard Contractual Clauses (SCCs)', *Nordic Journal of European Law*, vol. 4, no. 2, pp. 37-47.

Corrales Compagnucci, M 2011, Protecting Patients' Rights in Clinical Trial Scenarios: The 'Bee Metaphor' and the Symbiotic Relationship. in M Bottis (ed.), *An Information Law for the 21st Century*. Nomiki Bibliothiki Group, pp. 5-13.

Corrales Compagnucci, M, Forgó, N, Kono, T, Teramoto, S & Vermeulen, EPM (eds) 2020, *Legal Tech and the New Sharing Economy*. Perspectives in Law, Business and Innovation, Springer Nature Singapore, Singapore.

Corrales Compagnucci, M, Kono, T & Teramoto, S 2019, Legal Aspects of Decentralized and Platform-Driven Economies. in M Corrales Compagnucci, N Forgó, T Kono, S Teramoto & EPM Vermeulen (eds), *Legal Tech and the New Sharing Economy.* Springer, Springer,

Perspectives in Law, Business and Innovation, pp. 1-14. https://doi.org/10.1007/978-981-15-1350-3_1

Corrales Compagnucci, M 2019, *Big Data, Databases and "Ownership" Rights in the Cloud*. Perspectives in Law, Business and Innovation, Springer, Singapore. https://doi.org/10.1007/978-981-15-0349-8

Corrales Compagnucci, Marcelo and Fenwick, Mark and Haapio, Helena and Vermeulen, Erik P.M., Integrating Law, Technology and Design: Teaching Data Protection & Privacy Law in a Digital Age (June 29, 2021). Available at SSRN: https://ssrn.com/abstract=3876281 or http://dx.doi.org/10.2139/ssrn.3876281

Corrales Compagnucci, M, Minssen, T, Seitz, C & Aboy, M 2020, 'Lost on the High Seas without a Safe Harbor or a Shield? Navigating Cross-Border Data Transfers in the Pharmaceutical Sector After Schrems II Invalidation of the EU-US Privacy Shield', *European Pharmaceutical Law Review*, vol. 4, no. 3, pp. 153-160.

Corrales Compagnucci, M, Meszaros, J, Minssen, T, Arasilango , A, Ous , T & Rajarajan, M 2019, 'Homomorphic Encryption: The 'Holy Grail' for Big Data Analytics & Legal Compliance in the Pharmaceutical and Healthcare Sector?', *European Pharmaceutical Law Review*, vol. 3 , no. 4, pp. 144-155. https://doi.org/10.21552/eplr/2019/4/5

Dahi, Alan and Corrales Compagnucci, Marcelo, Smart Devices - One of Many Catalysts to Reconsider the Controller-Processor Relationship of the GDPR (July 5, 2020). Available at SSRN: https://ssrn.com/abstract=3643751 or http://dx.doi.org/10.2139/ssrn.3643751

Djemame, K, Barnitzke, B, Corrales Compagnucci, M, Kiram, M, Jiang, M, Armstrong, D, Forgó, N & Nwankwo, I 2013, 'Legal Issues in Clouds: Towards a Risk Inventory', *Philosophical Transactions of the Royal Society A: Mathematical, Physical and Engineering Sciences*, vol. 371, no. 1983, pp. 1-17. https://doi.org/10.1098/rsta.2012.0075

Dolley S, Big Data's Role in Precision Public Health, Frontiers in Public Health, March 2018 | Volume 6 | Article 68.

Dunn A and others, Social media interventions for precision public health: promises and risks, npj Digital Medicine (2018)1:47.

Fedeli, P, Scendoni, R, Cingolani, M, Corrales Compagnucci, M, Cirocchi, R & Cannovo, N 2022, 'Informed Consent and Protection of Personal Data in Genetic Research on COVID-19', *Healthcare*, vol. 10, no. 2. https://doi.org/10.3390/healthcare10020349

Ferrer, AJ, Hernández, F, Tordsson, J, Elmroth, E, Ali-Eldin, A, Zsigri, C, Sirvent, R, Guitart, J, Badia, R, Djemame, K, Ziegler, W, Dimitrakos, T, Nair, S, Kousiouris, G, Konstanteli, K, Varvarigou, T, Hudzia, B, Kipp, A, Wesner, S, Corrales Compagnucci, M, Forgó, N, Sharif, T & Sheridan, C 2011, 'OPTIMIS: A Holistic Approach to Cloud Service Provisioning', *Future Generation Computer Systems - The International Journal of eScience*, vol. 28, no. 1, pp. 66.

Gerke S, Young S and Cohen G, Ethical and Legal Aspects of Ambient Intelligence in Hospitals, *JAMA* (Jan. 24, 2020).

Guidelines 03/2020 on the processing of data concerning health for the purpose of scientific research in the context of COVID-19, available at: https://edpb.europa.eu/our-work-tools/our-documents/ohjeet/guidelines-032020-processing-data-concerning-health-purpose_en Accessed 20 Jan 2021.

Jayanthi P and others, 'Fourth Industrial Revolution: An Impact on Healthcare Industry', p. 58. In: Tareq AHRAM (ed) *Advances in Artificial Intelligence, Software and Systems Engineering* (Springer, 2019).

Jurcys, Paulius and Corrales Compagnucci, Marcelo and Fenwick, Mark, The Future of International Data Transfers: Managing New Legal Risk with a 'User-Held' Data Model (January 17, 2022). Available at SSRN: https://ssrn.com/abstract=4010356 or http://dx.doi.org/10.2139/ssrn.4010356

Kiran M., Khan A.U., Jiang M., Djemame K., Oriol M. and Corrales, M (2012) Managing security threats in Clouds. *Digital Research.*

Kirkham, T, Armstrong, D, Djemame, K, Corrales Compagnucci, M, Kiran, M, Nwankwo, I, Jiang, M & Forgó, N 2012, Assuring Data Privacy in Cloud Transformations. in *Proceedings of the 11th IEEE International Conference on Trust, Security and Privacy in Computing and Communications (IEEE TrustCom-12)*. IEEE, pp. 1063. https://doi.org/10.1109/TrustCom.2012.97

Louwerse M and others, Social cues in animated conversational agents. Applied Cognitive Psychology: The Official Journal of the Society for Applied Research in Memory and Cognition, 19(6):693704, 2005.

Meszaros, J, Ho, C & Corrales Compagnucci, M 2020, Nudging Consent & the New Opt-Out System to the Processing of Health Data in England. in M Corrales Compagnucci, N Forgó, T Kono, S Teramoto & E Vermeulen (eds), *Legal Tech and the New Sharing Economy.* Springer, Singapore, Perspectives in Law, Business and Innovation, pp. 61-82. https://doi.org/10.1007/978-981-15-1350-3_5

Meszaros, Janos and Corrales Compagnucci, Marcelo and Minssen, Timo, The Interaction of the Medical Device Regulation and the GDPR: Do European Rules on Privacy and Scientific Research Impair the Safety & Performance of AI Medical Devices? (March 20, 2021). In: I. Glenn Cohen, Timo Minssen, W. Nicholson Price II, Christopher Robertson, and Carmel Shachar. The Future of Medical Device Regulation: Innovation and Protection, Cambridge University Press, Available at SSRN: https://ssrn.com/abstract=3808384

Minssen, T, Seitz, C, Aboy, M & Corrales Compagnucci, M 2020, 'The EU-US Privacy Shield Regime for Cross-Border Transfers of Personal Data under the GDPR: What are the legal challenges and how might these affect cloud-based technologies, big data, and AI in the medical sector?', *European Pharmaceutical Law Review*, vol. 4, no. 1, pp. 34 - 50. https://doi.org/10.21552/eplr/2020/1/6

Oliva, M and Corrales Compagnucci, M 2011, 'Law Meets Biology: Are Our Databases Eligible for Legal Protection?', *SCRIPTed: A Journal of Law, Technology & Society*, vol. 8, no. 3, pp. 226-228.

Pagallo U, Corrales M, Fenwick M, Forgó N (2018) The Rise of Robotics & AI: Technological Advances and Normative Dilemmas. In: Corrales M, Fenwick M, Forgó N (2018), *Robotics, AI and the Future of Law.* Springer, Singapore

Panesar A, *Machine Learning and AI for Healthcare: Big Data for Improved Health Outcomes* (Apress, 2019).

Prosperi M, Big data hurdles in precision medicine and precision public health, BMC Medical Informatics and Decision Making, 2018, 18:139.

Wachter R, Why Diagnostic Errors Don't Get Any Respect and What Can Be Done About Them. Health Affairs, Aug 2017.

Yang G (2017) Office Operating Problem Scoring System Based on AI in: Hui YANG (ed) *Artificial Intelligence: Science and Technology*, Proceedings of the 2016 International Conference (AIST 2016), Shanghai, China, World Scientific.